\documentclass{article} 
\usepackage{iclr2027_conference,times}
\usepackage{wrapfig}
\usepackage{booktabs,tabularx,CJKutf8}

\usepackage{amsmath,amsfonts,bm}

\def\eqref#1{equation~\ref{#1}}

\def\1{\bm{1}}

\DeclareMathAlphabet{\mathsfit}{\encodingdefault}{\sfdefault}{m}{sl}
\SetMathAlphabet{\mathsfit}{bold}{\encodingdefault}{\sfdefault}{bx}{n}

\usepackage{hyperref}
\usepackage{url}
\usepackage{booktabs}
\usepackage{graphicx}
\usepackage{capt-of}
\title{Collapse, Repair, and Drift: Dissecting Fine Tuning Dynamics in Speech Language Models}

\iclrfinalcopy

\title{Collapse, Repair, and Drift:
Dissecting Fine Tuning Dynamics in Speech Language Models}

\author{%
\textbf{Yichi Wang}$^{1,3,\dagger}$ \quad
\textbf{Wangjin Zhou}$^{1,3,\dagger,*}$ \quad
\textbf{Yizhou Zhang}$^{1,3}$ \quad
\textbf{Xin Gu}$^{2,3}$ \\[0.1em]
\textbf{Wei Tan}$^{3}$ \quad
\textbf{Yi Zhao}$^{3}$ \quad
\textbf{Gong Zhi}$^{3}$ \\[0.2em]
$^{1}$Department of Intelligence Science and Technology, Kyoto University \\
$^{2}$Communication University of China \\
$^{3}$Xiaowei, WXG, Tencent \\[0.2em]
\texttt{\{yiwang,zhou\}@sap.ist.i.kyoto-u.ac.jp}
}

\begin{document}
\raggedbottom

\maketitle

\begingroup
\renewcommand{\thefootnote}{\fnsymbol{footnote}}
\footnotetext[2]{Equal contribution.}
\footnotetext[1]{Corresponding author.}
\endgroup

\lhead{Preprint}

\begin{abstract}
Supervised fine tuning (SFT) is commonly monitored through training and evaluation loss, but loss does not fully characterize how model functionality evolves along the same trajectory. We study this gap through dialect adaptation of Qwen3-ASR, MiMo-Audio, and FireRed-LLM across learning rates from $10^{-5}$ to $10^{-3}$. Character error rate (CER) reveals pronounced transient collapse and recovery whose severity and temporal structure are only weakly reflected by loss. Bidirectional layer patching reveals strong cross checkpoint functional incompatibility during recoverable collapse, with model dependent depth profiles that reorganize during recovery. Pairwise gradient similarity alone does not reliably indicate optimization stability, whereas raw gradient spectral analysis provides a clearer structural distinction between stable and unstable optimization: successfully converged runs approach a broad and temporally stable spectral energy distribution, while failed runs remain spectrally concentrated or exhibit persistent fluctuations. Finally, AdamW update transfer bypasses pronounced early collapse, yet substantial Mandarin degradation still emerges later in training. These results motivate three separable SFT dynamics: \emph{collapse}, \emph{repair}, and \emph{drift}, and show that transient collapse is not necessary for persistent source domain degradation.
\end{abstract}

\section{Introduction}

Supervised fine tuning (SFT) is widely used to adapt pretrained models to new tasks and domains \citep{zhang-etal-2023-speechgpt, fong25_interspeech, zhou2026rethinkingspeechfoundationmodel}. During training, optimization is typically monitored through training and evaluation loss, and abnormal behaviors such as loss spikes or plateaus are often used to diagnose changes in optimization dynamics \citep{tacl_a_00708, mircea-etal-2025-training, wang2025learning, ICLR2025_63e88a6e}. However, loss measures the optimized objective and does not directly reveal how the model's inference-time functionality evolves along the same trajectory \citep{pmlr-v202-liu23ao, xia-etal-2023-training}. This raises a basic question: what information about fine tuning dynamics is missed when training is characterized primarily through the loss trajectory?

Speech recognition provides a useful setting for studying this question because acoustic input and textual output have strong temporal and content correspondence. Unlike open-ended tasks such as captioning, recognition errors can be localized and quantified against a well-defined reference, making functional changes easier to attribute along the training trajectory. This allows supervised loss and character error rate (CER) to provide complementary optimization and behavioral views of the same SFT process \citep{pmlr-v202-radford23a, zhang2026directmediated}.

\begin{wrapfigure}{r}{0.48\textwidth}
    \centering
    \includegraphics[width=\linewidth]{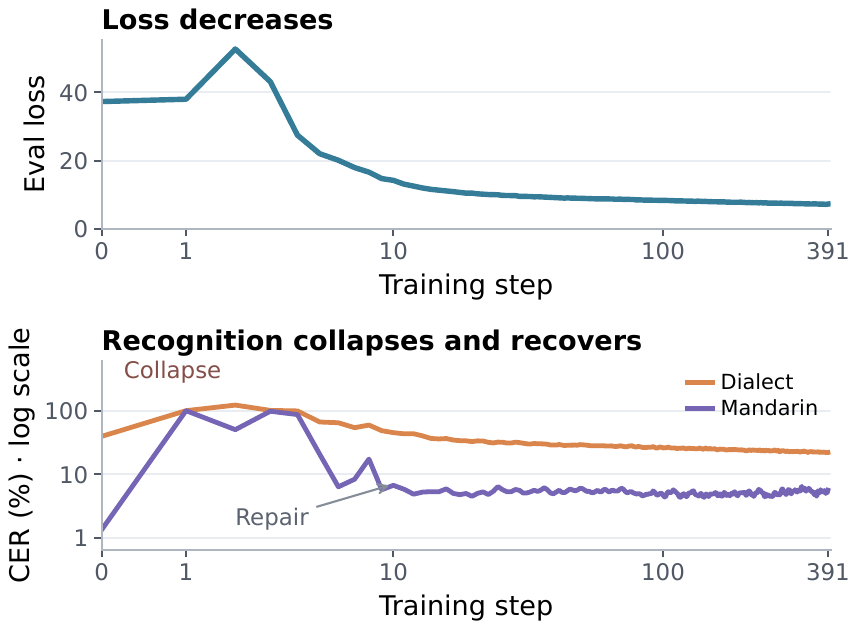}
    \caption{Evaluation loss and recognition CER can exhibit qualitatively different trajectories during SFT.}
    \label{fig:loss_cer_teaser}
\end{wrapfigure}

In our experiments, these two views can diverge substantially. As illustrated in Fig.~\ref{fig:loss_cer_teaser}, evaluation loss may vary only moderately while recognition performance deteriorates sharply. We observe this behavior when adapting Qwen3-ASR \citep{shi2026qwen3asr}, MiMo-Audio \citep{xiaomi2025mimoaudio}, and FireRed-LLM \citep{xu2025fireredasr} to newly collected dialect speech across learning rates from $10^{-5}$ to $10^{-3}$. Depending on the model and learning rate, CER ranges from smooth adaptation to pronounced transient collapse followed by recovery, and, at larger optimization scales, to persistent failure.

To examine the transient collapse, we use bidirectional layer patching between pretrained and early SFT checkpoints. The results reveal strong and structured cross checkpoint functional incompatibility during recoverable collapse, with model dependent depth profiles that reorganize as recognition recovers.

We then examine whether gradient dynamics provide a reliable description of optimization state. Successfully converged runs show characteristic within-step and inter-step cosine patterns, but similar patterns can also appear transiently earlier in training, making pairwise gradient similarity an unreliable stability indicator. In contrast, the singular value spectrum of the raw gradient matrix provides a clearer structural distinction between stable and unstable optimization. Successfully converged runs approach a broad and temporally stable spectral energy distribution, whereas failed runs remain spectrally concentrated or exhibit persistent spectral fluctuations.

Finally, we test whether early collapse is necessary for the persistent source domain degradation observed later in adaptation. Prior work has shown that abrupt performance degradation can remain recoverable \citep{zheng2025spurious}. Using AdamW update transfer, we modify the early optimization path in Qwen3-ASR at $10^{-4}$ and $10^{-5}$ and FireRed-LLM at $10^{-4}$. In the two $10^{-4}$ settings, the intervention bypasses pronounced early collapse while gradually returning the student to ordinary SFT. Mandarin CER nevertheless subsequently drifts toward the degraded teacher regime, showing that transient collapse is not necessary for persistent source domain degradation.

Taken together, these observations reveal three separable dynamics during speech model SFT: rapid \emph{collapse}, subsequent \emph{repair}, and slower source domain \emph{drift}.

Our main contributions are:
\begin{itemize}
    \item We show that severe recognition changes can be only weakly reflected in the loss trajectory, exposing transient SFT dynamics that loss alone does not characterize well.
    \item We identify structured cross checkpoint functional incompatibility during recoverable collapse and show that raw gradient spectral structure provides a clearer distinction between stable and unstable optimization than pairwise gradient similarity alone.
    \item We show through AdamW update transfer that bypassing pronounced early collapse does not prevent persistent source domain degradation, separating collapse from long term drift.
\end{itemize}

\section{Related Work}

\subsection{Speech Language Model Adaptation}

Modern speech language models connect acoustic representations to pretrained language models through speech encoders and adaptation modules. FireRed-LLM\citep{xu2025fireredasr}, Qwen3-ASR \citep{shi2026qwen3asr}, and MiMo-Audio \citep{xiaomi2025mimoaudio} provide independently developed systems with different architectures and training recipes . We use them as complementary settings for studying SFT dynamics across models.

\subsection{Transient Degradation and Forgetting}

Forgetting is commonly evaluated from model performance before and after adaptation, while recent work has begun to study the trajectory itself. Spurious forgetting shows that large early performance drops can arise even when the underlying capability remains recoverable \citep{ICLR2024_c4de749a, zheng2025spurious}. This motivates separating transient functional degradation from persistent capability loss, but their dynamical relationship remains unclear. We study both within the same SFT trajectory and directly intervene on the early optimization path to test whether collapse is necessary for later source domain drift.

\subsection{Functional and Optimization Diagnostics}

Prior studies characterize how speech and text representations are organized across LLM depth, including directional alignment and broad cross layer correspondence \citep{xiang2025modalitygap,hsu2026anatomy, zhang2026directmediated, glazer2026audiolanguagemodelslisteningaudiospecialist, glazer2026beyond}. These analyses focus on representation structure within trained models. We instead study functional compatibility across checkpoints, using bidirectional layer patching to localize how pretrained and adapted computations become incompatible during collapse and reorganize during recovery.

Gradient spectra provide a complementary view of optimization structure beyond scalar loss. Spectral Lens shows that singular spectra of per sample gradients can reveal optimization changes not apparent from loss alone \citep{gur-ari2019gradient, liu2026spectrallens}. We combine pairwise gradient similarity with spectral analysis to examine whether behavioral recovery coincides with stabilization of the spectral energy distribution.

Together, these lines of work motivate connecting transient behavior, functional compatibility, optimization geometry, and persistent forgetting within the same SFT trajectories.

\section{Experimental Setup}

\subsection{Models and Adaptation Data}
We study three independently developed LLM-based ASR models:
Qwen3-ASR-1.7B,
MiMo-Audio-7B-Instruct, and
FireRedASR2-LLM.
For all models, we freeze the acoustic encoder and audio-to-language
projector and fine-tune the text decoder.

We adapt all models on a self-collected private speech corpus
assembled for this study, covering five Chinese dialects:
Changsha, Hakka, Minnan, Nanchang, and Ningbo. As the corpus is private
and newly collected, it is separate from the publicly documented
training data of these models.
The training, validation, and test splits contain 250,000, 25,000,
and 5,000 utterances, respectively.

We evaluate target-domain adaptation using dialect CER and source-domain
retention using Mandarin CER on 1,683 utterances from SpeechIO-3~\citep{du2025speechcolab}.
Full dataset statistics and the decoding and scoring protocol are
provided in Appendix~\ref{app:general_setup}.

\subsection{Optimization Conditions}
For each model, we perform SFT at learning rates
$10^{-5}$, $10^{-4}$, and $10^{-3}$, while keeping all other
optimization settings fixed within the model. We use AdamW with a
constant learning rate and no warm-up in the main experiments, so that
early training dynamics are observed directly under the target
optimization scale. The effect of learning-rate warm-up is examined
separately in Appendix~\ref{app:warmup}.

Training runs for one full pass over the adaptation set, corresponding to $T=391$ optimizer updates. 
Each update uses a dialect-balanced effective batch of 640 utterances, with 128 utterances from each of the five
dialects. To characterize the training trajectory rather than only the endpoint, we track loss and recognition performance throughout SFT.
Detailed optimization and batching configurations are provided in
Appendix~\ref{app:models_training}.

\subsection{Gradient Measurements}
\label{sec:gradient_measurements}

Let $T=391$ denote the total number of optimizer updates and
$i\in\{1,\ldots,K\}$ index the $K=5$ dialects. For
$t=1,\ldots,T$, let $g_{t,i}$ denote the gradient computed from
dialect $i$ at the model state $\theta_{t-1}$, immediately before
optimizer update $t$. We normalize each gradient as
$\bar g_{t,i}=g_{t,i}/\lVert g_{t,i}\rVert_2$ and measure within-step
cross-dialect agreement by
\begin{equation}
S_{\mathrm{intra}}(t)
=
\frac{2}{K(K-1)}
\sum_{i<j}
\bar g_{t,i}^{\top}\bar g_{t,j}.
\end{equation}

For $t\geq2$, we measure temporal agreement within each dialect by
$r_{t,i}=\bar g_{t,i}^{\top}\bar g_{t-1,i}$ and summarize across
dialects as
\begin{equation}
S_{\mathrm{inter}}(t)
=
\operatorname{median}_{i=1,\ldots,K} r_{t,i}.
\end{equation}


For spectral analysis, we retain gradient magnitudes and construct
\[
G_{\mathrm{raw}}^{(t)}
= [g_{t,1},\ldots,g_{t,K}]^\top.
\]
Let $\sigma_k^{(t)}$ denote its singular values. We define the
normalized spectral energy and effective dimensionality as
\[
p_k^{(t)}
= \frac{(\sigma_k^{(t)})^2}
       {\sum_{j=1}^{K}(\sigma_j^{(t)})^2},
\qquad
D_{\mathrm{eff}}^{(t)}
= \frac{1}{\sum_{k=1}^{K}(p_k^{(t)})^2}.
\]
Further implementation details are provided in Appendices~\ref{app:gradient_similarity} and~\ref{app:gradient_spectrum}.

\section{Phenomenon: SFT Exhibits Distinct Dynamical Regimes}
\label{sec:phenomenon}

\begin{figure}[t]
    \centering
    \includegraphics[width=\linewidth]{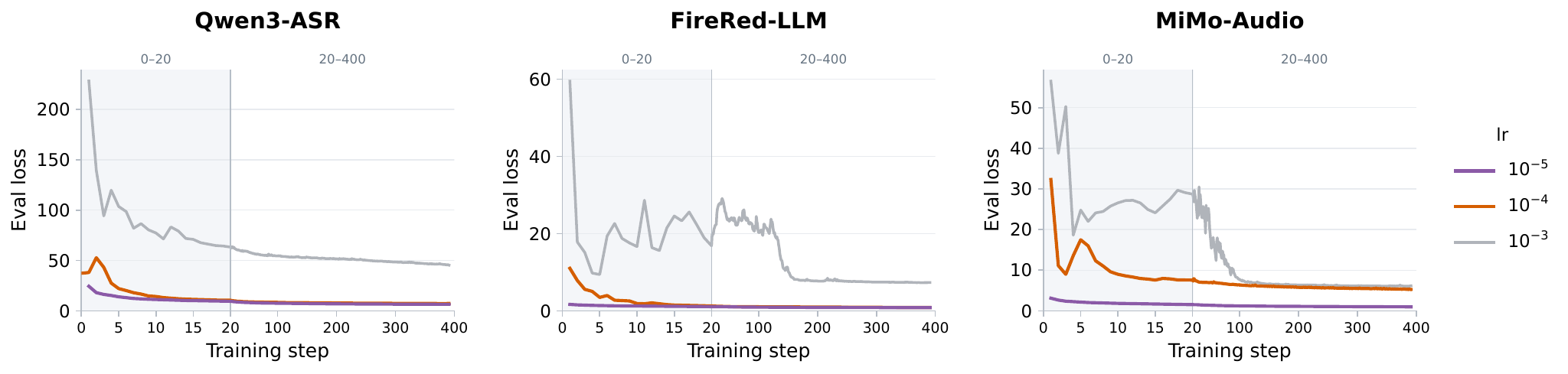}
    \caption{Evaluation loss trajectories during SFT across three models and learning rates.}
    \label{fig:eval_loss}
\end{figure}

\begin{figure}[b]
    \centering
    \includegraphics[width=\linewidth]{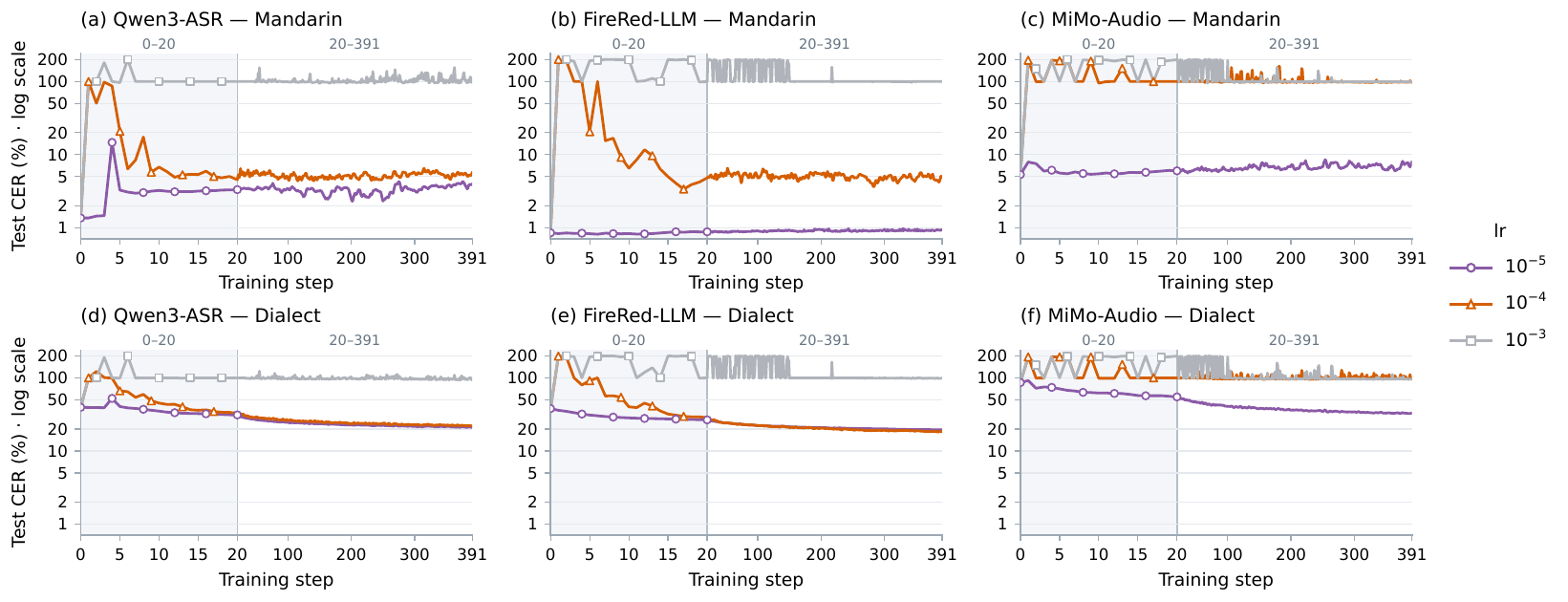}
    \caption{Test CER curves during supervised fine-tuning across three models and learning rates on Mandarin (top) and dialect (bottom) test sets. The piecewise-linear x-axis expands steps 0--20.}
    \label{fig:cer_curves}
\end{figure}

\subsection{Loss Curves Hide Recognition Dynamics}
As shown in Fig.~\ref{fig:eval_loss}, evaluation loss reflects the overall effect of learning rate on optimization, but provides a comparatively coarse view of the transient training trajectory. At $10^{-5}$, loss decreases rapidly for most runs. At $10^{-3}$, it increases early and then decreases, eventually converging to a substantially higher level than the lower learning rate runs. These curves therefore distinguish broadly successful and failed optimization regimes, but do not fully expose the severity or temporal structure of the intermediate functional changes.

CER provides a more direct view of model performance. As shown in Fig.~\ref{fig:cer_curves}, recognition can deteriorate sharply within the first few updates and then recover rapidly before entering ordinary adaptation. At $10^{-3}$, the same early collapse is followed by only partial recovery, leaving recognition substantially worse than the base model. Thus, loss and CER broadly reflect the same training regimes, but CER makes the severity and recovery process of recognition degradation more directly observable.

\subsection{Model Behavior at $10^{-5}$}

At a learning rate of $10^{-5}$, the three models occupy different dynamical regimes. FireRed-LLM exhibits the smoothest adaptation trajectory. Dialect CER decreases from the beginning of training and gradually slows as the model approaches convergence, while Mandarin CER remains nearly unchanged. This run therefore provides a reference case in which target-domain adaptation proceeds without observable functional collapse or substantial forgetting.

Qwen3-ASR and MiMo-Audio behave differently. Their dialect CER shows a mild early degradation before entering a normal convergence trajectory. Mandarin CER also deteriorates slightly at the beginning of training, then partially recovers before continuing along a gradual forgetting trajectory.

These trajectories already indicate that early functional collapse, target-domain learning, and source-domain forgetting are not the same process. FireRed-LLM adapts successfully without observable collapse, whereas Qwen3-ASR and MiMo-Audio undergo a mild collapse-and-repair cycle before final convergence, accompanied by persistent but relatively stable source-domain forgetting.

\subsection{Model Behavior at $10^{-4}$}

When the learning rate is increased to $10^{-4}$, both dialect and Mandarin CER rise sharply at the beginning of training across all three models. FireRed-LLM and Qwen3-ASR then recover rapidly and proceed toward convergence. Their final dialect CER is comparable to that obtained at $10^{-5}$, indicating that successful target-domain adaptation is still achieved. Source-domain retention is worse than at $10^{-5}$, however: after the initial recovery, Mandarin CER remains at a persistently degraded level. Viewed from the trajectory alone, this pattern can give the impression that the later forgetting originates from the severe disruption at the beginning of training.

MiMo-Audio behaves differently. After the initial CER explosion, it fails to recover into a normally convergent regime and remains substantially degraded. This behavior resembles the persistent failure observed for all three models at $10^{-3}$.

\subsection{Persistent Failure at $10^{-3}$}

At $10^{-3}$, all three models again exhibit a sharp initial increase in both dialect and Mandarin CER. Recognition performance then remains at a high level throughout training, failing to return to the base-model performance, let alone approach the normally adapted solutions obtained at lower learning rates. Evaluation loss behaves differently: after a brief early increase, it decreases rapidly, yet its final value remains more than twice that of the normally converged lower learning rate runs.

This regime is important because it distinguishes transient collapse from persistent failure. An early recovery response alone does not guarantee that the subsequent adaptation process will enter a stable training regime.

\section{Functional Localization: Collapse Induces Structured Functional Incompatibility}
\label{sec:functional_localization}

\subsection{Bidirectional Layer Patching}

We analyze transient functional changes during training through bidirectional layer patching between the pretrained model and early SFT checkpoints. For a model with $L$ LLM layers, at boundary $n$ we construct two hybrid models. The forward hybrid uses pretrained layers $1$ through $n$ followed by SFT layers $n+1$ through $L$, while the reverse hybrid uses the opposite assignment. Sweeping $n$ produces a depthwise CER profile that measures the functional compatibility between pretrained and adapted computation.

We focus on a sequence of closely spaced checkpoints around the onset, peak, and recovery of collapse. If replacing part of a degraded SFT model with pretrained layers restores recognition, the corresponding region contributes to the functional incompatibility. Reverse patching provides a complementary view by identifying where introducing SFT computation reintroduces the degraded behavior, thereby localizing the network regions associated with this incompatibility.

\subsection{Collapse and Repair}

\begin{figure}[t]
    \centering
    \includegraphics[width=\linewidth]{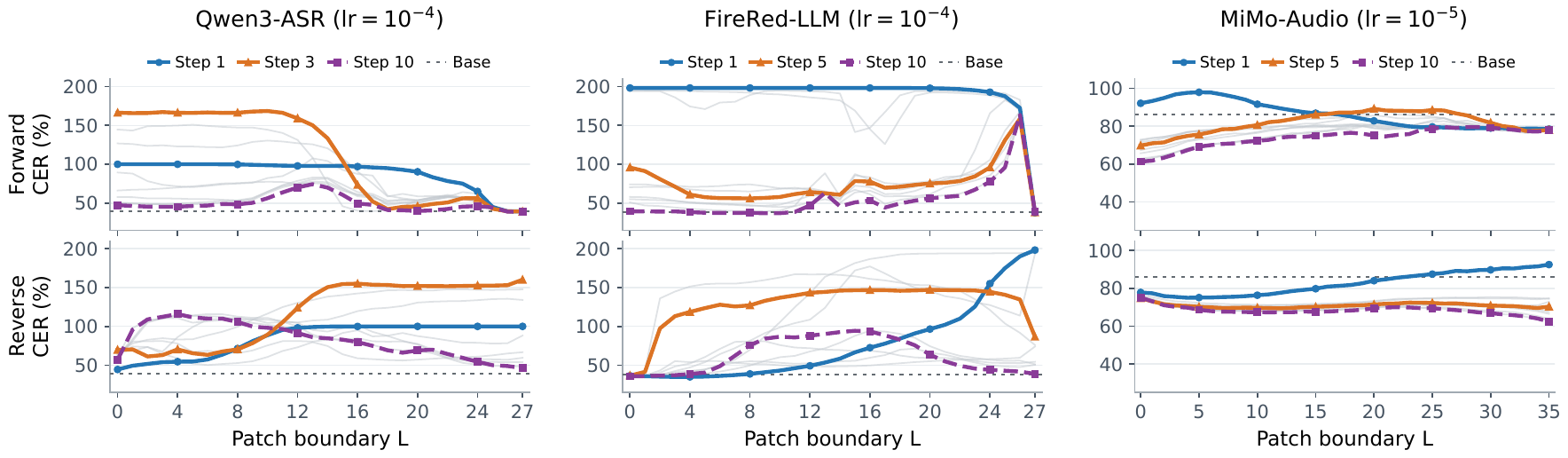}
    \caption{Bidirectional layer patching across models. Qwen3-ASR and FireRed-LLM use a learning rate of $10^{-4}$, while MiMo-Audio uses $10^{-5}$. The top and bottom rows show forward and reverse patching, respectively. Colored curves highlight selected checkpoints; gray curves show the remaining checkpoints. The dashed Base line in each column marks the corresponding pretrained model's decoding CER.}
    \label{fig:patching}
\end{figure}

As shown in Fig.~\ref{fig:patching}, all three models that undergo transient collapse and repair exhibit pronounced changes in their layer-patching profiles. However, the depthwise organization of these changes differs substantially across models.

The patching profiles show clear functional incompatibility between partially trained checkpoints and their corresponding pretrained models. The depth at which this incompatibility is strongest varies across both models and training steps, indicating that the disruption is not confined to a fixed layer region. Comparing these profiles with the CER trajectories further shows that the step-10 checkpoints, which lie in the recovery stage, generally exhibit smoother patching curves and lower overall mismatch with the pretrained computation than checkpoints near the collapse peak. These results indicate that recoverable early collapse is associated with a highly nonuniform disruption of computation across network depth, which becomes progressively reorganized during repair.

\section{Gradient Dynamics: Decorrelation and Spectral Stabilization}
\label{sec:gradient}

\subsection{Within-Step and Inter-Step Similarity Capture Complementary Dynamics}

\begin{figure}[t]
    \centering
    \includegraphics[width=\linewidth]{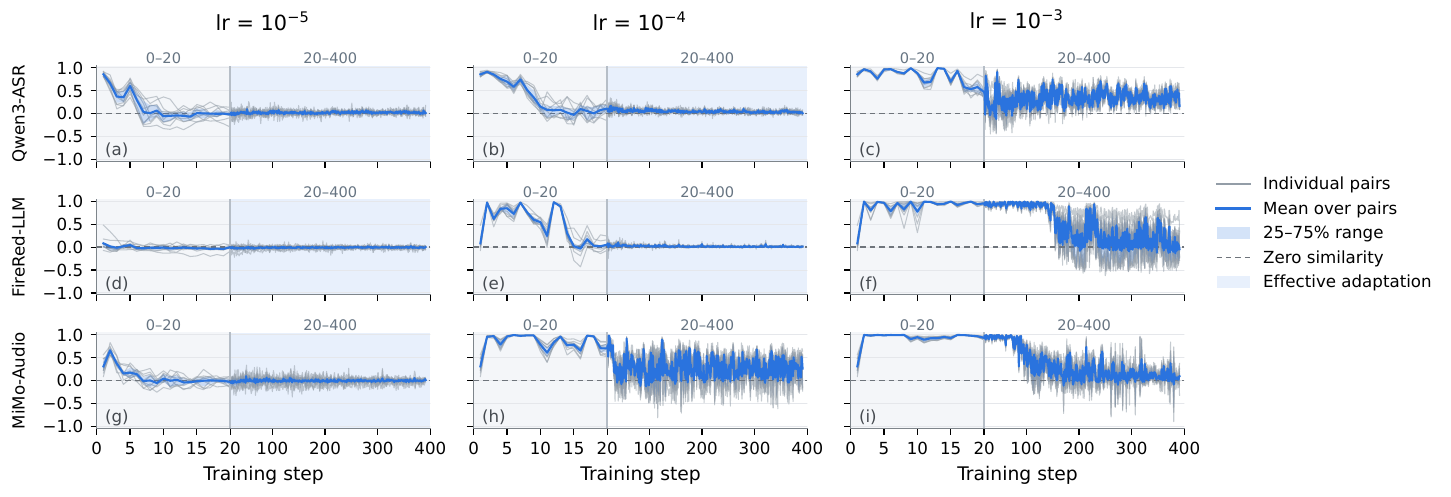}
    \par\vspace{2mm}
    \includegraphics[width=\linewidth]{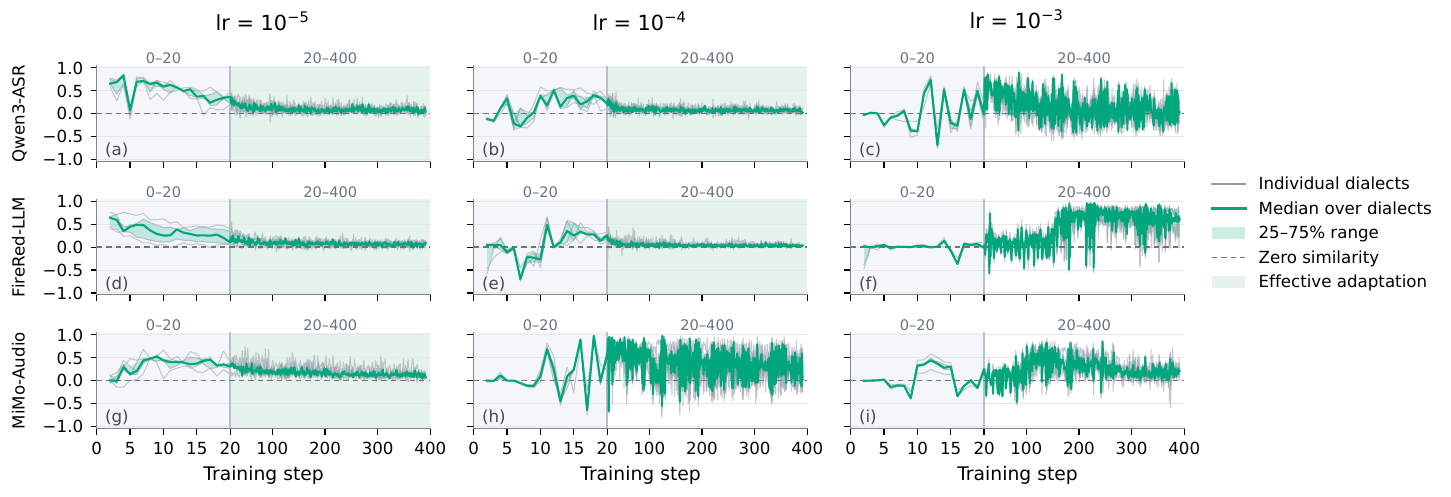}
    \caption{Gradient cosine similarity across models and learning rates.
    Top: within-step similarity across dialects.
    Bottom: similarity between adjacent steps within each dialect.
    Colored curves show the median, shaded bands indicate the
    interquartile range, and gray curves show individual series.}
    \label{fig:gradient_similarity}
\end{figure}

Figure~\ref{fig:gradient_similarity} summarizes two complementary forms of gradient similarity. $S_{\mathrm{intra}}$ measures the similarity between gradients from different dialects at the same optimization step, while $S_{\mathrm{inter}}$ measures the gradient similarity of the same dialect across adjacent training steps.

Under lower learning rates, runs that converge successfully generally exhibit rapid decorrelation. $S_{\mathrm{intra}}$ decreases quickly from relatively high initial similarity and then fluctuates within a narrow range around zero. As the learning rate increases, the gradients either retain stronger collinearity or exhibit substantially larger fluctuations, with the clearest deviations appearing in the failed runs at $10^{-3}$.

The inter-step similarity provides a complementary view. High $S_{\mathrm{inter}}$ can appear during early adaptation as well as in some failed high-learning-rate runs, while low $S_{\mathrm{inter}}$ can appear both transiently in early training and after successful convergence. Therefore, neither high nor low inter-step similarity alone provides a reliable indicator of convergence.

Considering both within-step and inter-step similarity, successfully converged runs can be qualitatively characterized by both statistics fluctuating within a narrow range around zero. However, the gradients lie in an extremely high-dimensional parameter space, where near-orthogonality can arise naturally and cosine similarity around zero is therefore only weakly diagnostic of the underlying optimization structure. To obtain a more compact structural view, we analyze the gradient matrix through singular value decomposition and examine how its energy is distributed across the resulting singular components.

\subsection{Spectral Stability Distinguishes Stable Adaptation from Unstable Optimization}

\begin{figure}[h]
    \centering
    \includegraphics[width=\linewidth]{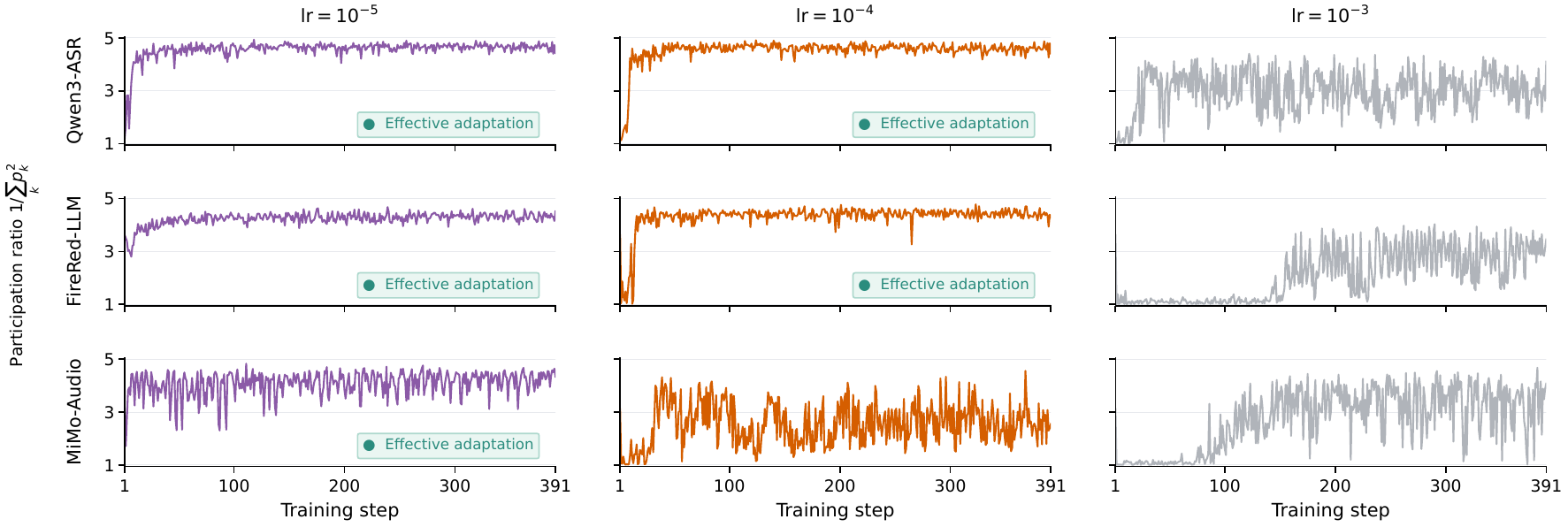}
    \caption{Gradient effective dimensionality across models and
    learning rates, measured by $D_{\mathrm{eff}}$.
    Values near 1 indicate concentrated spectral energy;
    values near 5 indicate approximately equal energy shares.}
    \label{fig:gradient_participation_lines}
\end{figure}

Figure~\ref{fig:gradient_participation_lines} shows how
$D_{\mathrm{eff}}^{(t)}$ evolves during training.
With five sampled gradients, $1\leq D_{\mathrm{eff}}^{(t)}\leq5$. Lower values indicate stronger spectral concentration, whereas values close to 5 indicate that the energy is distributed nearly uniformly across the five spectral components.
For runs that converge successfully, the effective dimensionality rises rapidly during early training and then remains close to 5 for an extended period, with only minor fluctuations. In contrast, failed runs either remain highly spectrally concentrated for extended periods or exhibit persistent large fluctuations in effective dimensionality throughout training.

These trajectories provide a clearer structural distinction between stable and unstable optimization than pairwise cosine similarity alone. Successful adaptation is associated with a nearly uniform and temporally stable spectral energy distribution, whereas failed optimization either remains dominated by a small number of spectral components or continues to undergo substantial spectral reorganization. The corresponding full singular-value energy distributions are provided in Appendix~\ref{app:spectral}.

\section{Bypassing Early Collapse Does Not Prevent Persistent Forgetting}
\label{sec:intervention}

\subsection{AdamW Update Transfer Changes Early Recognition Dynamics}

We examine whether source-domain degradation still emerges when the early optimization trajectory is modified to bypass transient recognition collapse. The intervention is applied to Qwen3-ASR at learning rates of $10^{-4}$ and $10^{-5}$ and to FireRed-LLM at $10^{-4}$. Qwen3-ASR and FireRed-LLM at $10^{-4}$ both exhibit pronounced early recognition collapse on the Mandarin and dialect test sets, whereas Qwen3-ASR at $10^{-5}$ shows only a smaller early Mandarin CER peak.

The student update is defined as
\begin{equation}
\theta_{s+1}^{S}
=
\theta_s^{S}
+
a_s u_s^{S}
+
b_s u_s^{T},
\end{equation}
where $u_s^{S}$ and $u_s^{T}$ denote the complete AdamW parameter displacements computed by the student and teacher at step $s$, respectively. During the early stage of training, the student first follows a scaled teacher update. The teacher contribution is then gradually reduced as the student's own updates progressively take over. Ordinary SFT is restored after $s=70$. Full implementation details and the coefficient schedule are provided in Appendix~\ref{app:update_transfer}.

\begin{figure}[h]
\centering
\includegraphics[width=\linewidth]{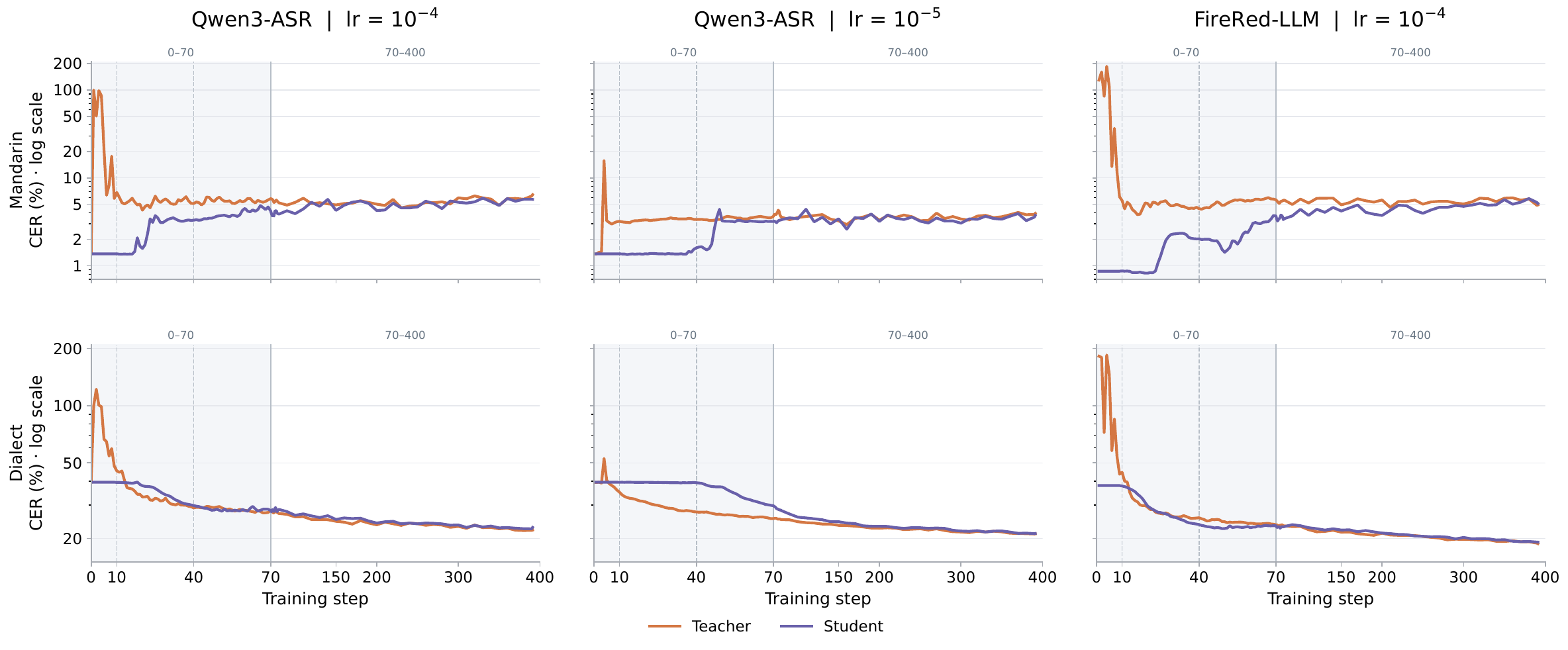}
\caption{CER trajectories of teacher and student models during SFT for Qwen3-ASR at $10^{-4}$ and $10^{-5}$ and FireRed-LLM at $10^{-4}$. The top row shows Mandarin speech and the bottom row shows dialect speech. The horizontal axis uses zero-indexed global batch steps: the student remains at the pretrained parameters during steps 0--9 and receives AdamW update transfer from step 10 to step 70.}
\label{fig:adamw_transfer}
\end{figure}

As shown in Fig.~\ref{fig:adamw_transfer}, update transfer substantially changes the early recognition dynamics. The student bypasses the pronounced Mandarin and dialect CER peaks observed in the teacher trajectories and subsequently moves gradually toward the teacher trajectory.

\subsection{Long-Term Drift Persists Without Early Collapse}

Despite bypassing the early collapse, the student still exhibits source-domain degradation during subsequent training. In the teacher trajectories, Mandarin CER rises sharply, recovers rapidly, and eventually enters an approximately stable plateau. This apparent plateau can be understood as the result of continued repair of the initial collapse occurring concurrently with progressive forgetting, with the two processes partially offsetting each other.

The intervention reduces the initial damage caused by collapse and therefore also reduces the amount of subsequent repair required. As the student's own AdamW updates progressively take over, Mandarin CER rises again while dialect performance continues to improve. This exposes a long-term drift that is less masked by concurrent recovery.

These results therefore separate early collapse from persistent forgetting: bypassing the former changes the temporal manifestation of source-domain degradation but does not eliminate the latter.

\section{Conclusion}

We studied complete SFT trajectories of Qwen3-ASR, MiMo-Audio, and FireRed-LLM across learning rates from $10^{-5}$ to $10^{-3}$. The resulting dynamics can be organized into three separable processes: \emph{collapse}, \emph{repair}, and \emph{drift}. Collapse corresponds to the rapid loss of recognition functionality during early adaptation, repair to the subsequent recovery from this disruption, and drift to the longer term degradation of source domain performance under continued adaptation. Their expression is strongly model and learning rate dependent: lower learning rates produce behaviors ranging from smooth adaptation to transient collapse, while larger learning rates increasingly lead to persistent recognition failure and unstable optimization.

These processes are reflected at multiple levels of the training dynamics. Bidirectional layer patching associates recoverable collapse with strong cross checkpoint functional incompatibility that weakens and reorganizes during repair, with model dependent depth profiles. Gradient similarity reveals substantial changes in within-step and inter-step organization during training, but near zero cosine similarity alone is weakly diagnostic in the high dimensional parameter space. The singular value spectrum provides a more compact structural view: successfully converged runs tend toward a broad and temporally stable spectral energy distribution, whereas failed runs remain spectrally concentrated or undergo persistent spectral reorganization. Behavioral recovery can therefore occur before the gradient spectrum reaches its later stable regime.

The AdamW update transfer intervention further separates collapse from long term drift. Bypassing the early recognition collapse does not eliminate subsequent source domain degradation, indicating that persistent forgetting can emerge without the initial collapse. The intervention also suggests that the apparently stable source domain CER observed after recovery can reflect continued repair of the initial disruption occurring alongside progressive forgetting. Reducing the initial collapse reduces the amount of subsequent repair and makes the longer term drift more directly observable.

Taken together, our results show that transient collapse, subsequent repair, and long term drift should be treated as distinct but interacting dynamical processes. Loss alone can obscure severe transient functional changes, pairwise gradient similarity does not uniquely characterize optimization stability, and behavioral recovery does not imply that the underlying optimization geometry has stabilized. Most importantly, the presence or absence of an early recognition collapse does not determine whether persistent source domain degradation will emerge during continued adaptation.

\section*{AI Use Statement}
We used ChatGPT to assist with English writing and language polishing,
and to help discover potentially relevant literature. The authors
independently checked the cited sources, reviewed all AI-assisted text,
and take full responsibility for the paper's content.

\bibliography{iclr2027_conference}

@inproceedings{zhang-etal-2023-speechgpt,
    title = "{S}peech{GPT}: Empowering Large Language Models with Intrinsic Cross-Modal Conversational Abilities",
    author = "Zhang, Dong  and
      Li, Shimin  and
      Zhang, Xin  and
      Zhan, Jun  and
      Wang, Pengyu  and
      Zhou, Yaqian  and
      Qiu, Xipeng",
    editor = "Bouamor, Houda  and
      Pino, Juan  and
      Bali, Kalika",
    booktitle = "Findings of the Association for Computational Linguistics: EMNLP 2023",
    month = dec,
    year = "2023",
    address = "Singapore",
    publisher = "Association for Computational Linguistics",
    url = "https://aclanthology.org/2023.findings-emnlp.1055/",
    doi = "10.18653/v1/2023.findings-emnlp.1055",
    pages = "15757--15773"
}

@inproceedings{fong25_interspeech,
  title     = {{Speech LLMs in Low-Resource Scenarios: Data Volume Requirements and the Impact of Pretraining on High-Resource Languages}},
  author    = {Seraphina Fong and Marco Matassoni and Alessio Brutti},
  year      = {2025},
  booktitle = {{Interspeech 2025}},
  pages     = {2003--2007},
  doi       = {10.21437/Interspeech.2025-764},
  issn      = {2958-1796},
}

@misc{zhou2026rethinkingspeechfoundationmodel,
      title={Rethinking Speech Foundation Model Fine-tuning: Better SFT or Better Match?}, 
      author={Wangjin Zhou and Yizhou Zhang and Yichi Wang and Tatsuya Kawahara},
      year={2026},
      eprint={2607.13864},
      archivePrefix={arXiv},
      primaryClass={cs.SD},
      url={https://arxiv.org/abs/2607.13864}, 
}

@article{tacl_a_00708,
    author = {Chang, Tyler A. and Tu, Zhuowen and Bergen, Benjamin K.},
    title = {Characterizing Learning Curves During Language Model Pre-Training: Learning, Forgetting, and Stability},
    journal = {Transactions of the Association for Computational Linguistics},
    volume = {12},
    pages = {1346-1362},
    year = {2024},
    month = {11},
    issn = {2307-387X},
    doi = {10.1162/tacl_a_00708},
    url = {https://doi.org/10.1162/tacl_a_00708},
    eprint = {https://direct.mit.edu/tacl/article-pdf/doi/10.1162/tacl_a_00708/2478609/tacl_a_00708.pdf},
}

@inproceedings{mircea-etal-2025-training,
    title = "Training Dynamics Underlying Language Model Scaling Laws: Loss Deceleration and Zero-Sum Learning",
    author = "Mircea, Andrei  and
      Chakraborty, Supriyo  and
      Chitsazan, Nima  and
      Rish, Irina  and
      Lobacheva, Ekaterina",
    editor = "Che, Wanxiang  and
      Nabende, Joyce  and
      Shutova, Ekaterina  and
      Pilehvar, Mohammad Taher",
    booktitle = "Proceedings of the 63rd Annual Meeting of the Association for Computational Linguistics (Volume 1: Long Papers)",
    month = jul,
    year = "2025",
    address = "Vienna, Austria",
    publisher = "Association for Computational Linguistics",
    url = "https://aclanthology.org/2025.acl-long.1366/",
    doi = "10.18653/v1/2025.acl-long.1366",
    pages = "28154--28188",
    ISBN = "979-8-89176-251-0"
}

@inproceedings{wang2025learning,
title={Learning Dynamics in Continual Pre-Training for Large Language Models},
author={Xingjin Wang and Howe Tissue and Lu Wang and Linjing Li and Daniel Dajun Zeng},
booktitle={Forty-second International Conference on Machine Learning},
year={2025},
url={https://openreview.net/forum?id=Vk1rNMl0J1}
}

@inproceedings{ICLR2025_63e88a6e,
 author = {Luo, Kairong and Wen, Haodong and Hu, Shengding and Sun, Zhenbo and Sun, Maosong and Liu, Zhiyuan and Lyu, Kaifeng and Chen, Wenguang},
 booktitle = {International Conference on Learning Representations},
 editor = {Y. Yue and A. Garg and N. Peng and F. Sha and R. Yu},
 pages = {39973--40017},
 title = {A Multi-Power Law for Loss Curve Prediction Across Learning Rate Schedules},
 url = {https://proceedings.iclr.cc/paper_files/paper/2025/file/63e88a6ebfb1f68808a058a32b947e71-Paper-Conference.pdf},
 volume = {2025},
 year = {2025}
}

@InProceedings{pmlr-v202-liu23ao,
  title = 	 {Same Pre-training Loss, Better Downstream: Implicit Bias Matters for Language Models},
  author =       {Liu, Hong and Xie, Sang Michael and Li, Zhiyuan and Ma, Tengyu},
  booktitle = 	 {Proceedings of the 40th International Conference on Machine Learning},
  pages = 	 {22188--22214},
  year = 	 {2023},
  editor = 	 {Krause, Andreas and Brunskill, Emma and Cho, Kyunghyun and Engelhardt, Barbara and Sabato, Sivan and Scarlett, Jonathan},
  volume = 	 {202},
  series = 	 {Proceedings of Machine Learning Research},
  month = 	 {23--29 Jul},
  publisher =    {PMLR},
  url = 	 {https://proceedings.mlr.press/v202/liu23ao.html}
}

@inproceedings{xia-etal-2023-training,
    title = "Training Trajectories of Language Models Across Scales",
    author = "Xia, Mengzhou  and
      Artetxe, Mikel  and
      Zhou, Chunting  and
      Lin, Xi Victoria  and
      Pasunuru, Ramakanth  and
      Chen, Danqi  and
      Zettlemoyer, Luke  and
      Stoyanov, Veselin",
    editor = "Rogers, Anna  and
      Boyd-Graber, Jordan  and
      Okazaki, Naoaki",
    booktitle = "Proceedings of the 61st Annual Meeting of the Association for Computational Linguistics (Volume 1: Long Papers)",
    month = jul,
    year = "2023",
    address = "Toronto, Canada",
    publisher = "Association for Computational Linguistics",
    url = "https://aclanthology.org/2023.acl-long.767/",
    doi = "10.18653/v1/2023.acl-long.767",
    pages = "13711--13738"
}

@InProceedings{pmlr-v202-radford23a,
  title = 	 {Robust Speech Recognition via Large-Scale Weak Supervision},
  author =       {Radford, Alec and Kim, Jong Wook and Xu, Tao and Brockman, Greg and Mcleavey, Christine and Sutskever, Ilya},
  booktitle = 	 {Proceedings of the 40th International Conference on Machine Learning},
  pages = 	 {28492--28518},
  year = 	 {2023},
  editor = 	 {Krause, Andreas and Brunskill, Emma and Cho, Kyunghyun and Engelhardt, Barbara and Sabato, Sivan and Scarlett, Jonathan},
  volume = 	 {202},
  series = 	 {Proceedings of Machine Learning Research},
  month = 	 {23--29 Jul},
  publisher =    {PMLR},
  url = 	 {https://proceedings.mlr.press/v202/radford23a.html}
}

@article{zhang2026directmediated,
title={Direct or Mediated? Task-Dependent Audio Information Routing in Large Audio Language Models},
author={Zhang, Yizhou and Zhou, Wangjin and Gu, Xin and Wang, Yichi and Tan, Wei and Zhao, Yi and Gong, Zhi and Imoto, Keisuke and Kawahara, Tatsuya},
journal={arXiv preprint arXiv:2608.27026},
year={2026}
}

@article{du2025speechcolab,
  title={SpeechColab leaderboard: An open-source platform for automatic speech recognition evaluation},
  author={Du, Jiayu and Li, Jinpeng and Chen, Guoguo and Zhang, Wei-Qiang},
  journal={Computer Speech \& Language},
  volume={94},
  pages={101805},
  year={2025},
  publisher={Elsevier},
  doi={10.1016/j.csl.2025.101805}
}

@article{xu2025fireredasr,
  title={FireRedASR: Open-Source Industrial-Grade Mandarin Speech Recognition Models from Encoder-Decoder to LLM Integration},
  author={Xu, Kai-Tuo and Xie, Feng-Long and Tang, Xu and Hu, Yao},
  journal={arXiv preprint arXiv:2501.14350},
  year={2025},
  url={https://arxiv.org/abs/2501.14350}
}

@article{shi2026qwen3asr,
  title={Qwen3-ASR Technical Report},
  author={Shi, Xian and Wang, Xiong and Guo, Zhifang and Wang, Yongqi and Zhang, Pei and Zhang, Xinyu and Guo, Zishan and Hao, Hongkun and Xi, Yu and Yang, Baosong and Xu, Jin and Zhou, Jingren and Lin, Junyang},
  journal={arXiv preprint arXiv:2601.21337},
  year={2026},
  url={https://arxiv.org/abs/2601.21337}
}

@article{xiaomi2025mimoaudio,
  title={MiMo-Audio: Audio Language Models are Few-Shot Learners},
  author={{Xiaomi LLM-Core Team}},
  journal={arXiv preprint arXiv:2512.23808},
  year={2025},
  url={https://arxiv.org/abs/2512.23808}
}

@inproceedings{xiang2025modalitygap,
  title={Understanding the Modality Gap: An Empirical Study on the Speech-Text Alignment Mechanism of Large Speech Language Models},
  author={Xiang, Bajian and Zhao, Shuaijiang and Guo, Tingwei and Zou, Wei},
  booktitle={Proceedings of the 2025 Conference on Empirical Methods in Natural Language Processing},
  pages={5187--5202},
  publisher={Association for Computational Linguistics},
  year={2025},
  doi={10.18653/v1/2025.emnlp-main.262},
  url={https://aclanthology.org/2025.emnlp-main.262/}
}

@article{hsu2026anatomy,
  title={Anatomy of the Modality Gap: Dissecting the Internal States of End-to-End Speech LLMs},
  author={Hsu, Ming-Hao and Zhang, Xueyao and Tian, Xiaohai and Zhang, Jun and Wu, Zhizheng},
  journal={arXiv preprint arXiv:2603.01502},
  year={2026},
  url={https://arxiv.org/abs/2603.01502}
}

@inproceedings{zheng2025spurious,
  title={Spurious Forgetting in Continual Learning of Language Models},
  author={Zheng, Junhao and Cai, Xidi and Qiu, Shengjie and Ma, Qianli},
  booktitle={International Conference on Learning Representations},
  year={2025},
  url={https://proceedings.iclr.cc/paper_files/paper/2025/hash/a774503daed55eb53c634847ae071ec7-Abstract-Conference.html}
}

@inproceedings{ICLR2024_c4de749a,
 author = {Zhang, Xiao and Wu, Ji},
 booktitle = {International Conference on Learning Representations},
 editor = {B. Kim and Y. Yue and S. Chaudhuri and K. Fragkiadaki and M. Khan and Y. Sun},
 pages = {45235--45255},
 title = {Dissecting learning and forgetting in language model finetuning},
 url = {https://proceedings.iclr.cc/paper_files/paper/2024/file/c4de749a4bd8802f0b4033d09a7867db-Paper-Conference.pdf},
 volume = {2024},
 year = {2024}
}

@misc{
gur-ari2019gradient,
title={Gradient Descent Happens in a Tiny Subspace},
author={Guy Gur-Ari and Daniel A. Roberts and Ethan Dyer},
year={2019},
url={https://openreview.net/forum?id=ByeTHsAqtX},
}

@article{liu2026spectrallens,
  title={Spectral Lens: Activation and Gradient Spectra as Diagnostics of LLM Optimization},
  author={Liu, Andy Zeyi and Paquette, Elliot and Sous, John},
  journal={arXiv preprint arXiv:2605.05683},
  year={2026},
  url={https://arxiv.org/abs/2605.05683}
}

@misc{glazer2026audiolanguagemodelslisteningaudiospecialist,
      title={Are Audio-Language Models Listening? Audio-Specialist Heads for Adaptive Audio Steering}, 
      author={Neta Glazer and Lenny Aharon and Ethan Fetaya},
      year={2026},
      eprint={2603.06854},
      archivePrefix={arXiv},
      primaryClass={cs.SD},
      url={https://arxiv.org/abs/2603.06854}, 
}

@inproceedings{glazer2026beyond,
  title     = {Beyond Transcription: Mechanistic Interpretability in ASR},
  author    = {Glazer, Neta and Segal-Feldman, Yael and Segev, Hilit and Shamsian, Aviv and Buchnick, Asaf and Hetz, Gill and Fetaya, Ethan and Keshet, Joseph and Navon, Aviv},
  booktitle = {Proceedings of the AAAI Conference on Artificial Intelligence},
  volume    = {40},
  number    = {44},
  pages     = {37407--37416},
  year      = {2026},
  doi       = {10.1609/aaai.v40i44.41073},
  url       = {https://ojs.aaai.org/index.php/AAAI/article/view/41073}
}
\bibliographystyle{iclr2027_conference}

\newpage
\appendix

\section{General Experimental Setup}
\label{app:general_setup}

\subsection{Datasets}
\label{app:datasets}

We use a self-collected private speech corpus covering five Chinese
dialects: Changsha, Hakka, Minnan, Nanchang, and Ningbo.
The dialect training, validation, and test sets contain
250,000, 25,000, and 5,000 utterances, respectively.
Mandarin recognition is evaluated separately on 1,683 utterances
from SpeechIO-3. Table~\ref{tab:dataset_statistics} summarizes
the number of utterances and audio duration for each split.

\begin{table}[h]
    \centering
    \caption{Dataset statistics. Duration is reported in hours.
    Mandarin data are from SpeechIO-3 and used only for evaluation.}
    \label{tab:dataset_statistics}
    \small
    \setlength{\tabcolsep}{6pt}
    \begin{tabular}{lrrrrrr}
        \toprule
        & \multicolumn{2}{c}{Training}
        & \multicolumn{2}{c}{Validation}
        & \multicolumn{2}{c}{Test} \\
        \cmidrule(lr){2-3}
        \cmidrule(lr){4-5}
        \cmidrule(lr){6-7}
        Dialect / Language
        & Utterances & Hours
        & Utterances & Hours
        & Utterances & Hours \\
        \midrule
        Changsha & 50,000 & 61.52
                 & 5,000 & 5.90
                 & 1,000 & 1.17 \\
        Hakka    & 50,000 & 69.82
                 & 5,000 & 7.03
                 & 1,000 & 1.40 \\
        Minnan   & 50,000 & 81.47
                 & 5,000 & 8.16
                 & 1,000 & 1.65 \\
        Nanchang & 50,000 & 38.36
                 & 5,000 & 3.88
                 & 1,000 & 0.78 \\
        Ningbo   & 50,000 & 136.02
                 & 5,000 & 13.66
                 & 1,000 & 2.67 \\
        \midrule
        Dialect total
                 & 250,000 & 387.18
                 & 25,000 & 38.62
                 & 5,000 & 7.66 \\
        \midrule
        Mandarin & -- & --
                 & -- & --
                 & 1,683 & 2.39 \\
        \bottomrule
    \end{tabular}
\end{table}

\subsection{Models and Training Details}
\label{app:models_training}

We study three models: Qwen3-ASR, MiMo-Audio, and FireRed-LLM.
For all three, we freeze the speech encoder and audio-to-language
projector and fine-tune the text decoder using paired audio and
reference transcripts. Table~\ref{tab:asr_templates} summarizes
the input templates and supervised tokens.

\begin{table*}[b]
\centering
\caption{Input templates and supervised tokens for the three ASR models.
Fixed input-side tokens and assistant prefixes are excluded from the loss.}
\label{tab:asr_templates}
\begin{tabularx}{\textwidth}{@{}lXX@{}}
\toprule
Model & Input and assistant prefix & Supervised tokens \\
\midrule
Qwen3-ASR
& Audio and the fixed prefix
  \texttt{language Chinese<asr\_text>}
& Transcript and EOS token \\
FireRed-LLM
& \texttt{<speech>} followed by
  \begin{CJK*}{UTF8}{gbsn}请转写音频为文字\end{CJK*}
& Plain-text transcript in the assistant response \\
MiMo-Audio
& Audio and the fixed instruction
  \begin{CJK*}{UTF8}{gbsn}请将这段语音转换为文字\end{CJK*};
  the assistant prefix includes an empty thinking block
& Plain-text transcript and \texttt{<|im\_end|>} \\
\bottomrule
\end{tabularx}
\end{table*}

Each model is independently initialized from its pretrained checkpoint
for learning rates $\eta\in\{10^{-3},10^{-4},10^{-5}\}$.
Other settings are fixed across learning rates within each model.

Each per-device micro-batch contains one utterance per dialect.
The effective batch size is
\[
B_{\mathrm{eff}}=B_{\mathrm{micro}}WA=5\times8\times16=640,
\]
where $B_{\mathrm{micro}}=5$ is the per-device micro-batch size,
$W=8$ is the number of GPUs, and $A=16$ is the number of
gradient-accumulation micro-steps. We define one training step as
one optimizer update after gradient accumulation; step 0 denotes
the pretrained checkpoint before any update. A full step therefore
uses 128 utterances per dialect. With 250,000 training utterances,
one pass through the training set requires
\[
\left\lceil\frac{250{,}000}{B_{\mathrm{eff}}}\right\rceil
=
\left\lceil\frac{250{,}000}{640}\right\rceil
=391
\]
training steps. We use AdamW with a constant learning rate and
no warm-up for these 391 steps. 

\subsection{Decoding and Evaluation}
\label{app:evaluation}

We generate transcripts autoregressively without teacher forcing,
setting the maximum generation length for each utterance to twice
the character length of its reference transcript. Before scoring,
we extract the transcript from each model's output and remove
model-specific control tokens and metadata. We then apply Unicode
NFKC normalization and lowercasing, and remove whitespace and
Unicode punctuation from both references and hypotheses.

For an utterance set $\mathcal{U}$, we compute corpus-level character
error rate (CER), expressed as a percentage, as
\[
\mathrm{CER}(\mathcal{U})
=100\frac{\sum_{u\in\mathcal{U}}(S_u+D_u+I_u)}
              {\sum_{u\in\mathcal{U}}N_u},
\]
where $S_u$, $D_u$, and $I_u$ denote the numbers of character
substitutions, deletions, and insertions under Levenshtein alignment,
and $N_u$ is the length of the normalized reference transcript.
Overall dialect CER is computed by pooling errors and reference
characters across all five dialects, rather than averaging
dialect-specific CERs. Mandarin CER is reported separately.

\section{Bidirectional Layer Patching}
\label{app:layer_patching}

\subsection{Hybrid Model Construction}
\label{app:patching_construction}

Following Section~\ref{sec:functional_localization}, let $M^{(0)}$
denote the pretrained model and $M^{(t)}$ the model after $t$
optimizer updates. Let $L$ be the number of Transformer layers
and $n$ the number of initial layers selected for the hybrid.
We construct two hybrids:
\[
\begin{aligned}
\widetilde M^{\mathrm{fwd}}_{t,n}
&=\bigl[M^{(0),1:n};M^{(t),n+1:L}\bigr],\\
\widetilde M^{\mathrm{rev}}_{t,n}
&=\bigl[M^{(t),1:n};M^{(0),n+1:L}\bigr].
\end{aligned}
\]
Both layer ranges are inclusive. In forward patching, layers
$1{:}n$ come from the pretrained checkpoint and layers
$(n+1){:}L$ from the checkpoint after $t$ SFT updates.
Reverse patching swaps these sources: layers $1{:}n$ come
from the SFT checkpoint and layers $(n+1){:}L$ from the
pretrained checkpoint.

We evaluate $n=1,\ldots,L$, where $L=28$ for Qwen3-ASR and
FireRed-LLM and $L=36$ for MiMo-Audio. The figures use a
zero-based layer boundary, so the plotted value is $n-1$.
Increasing this value selects more initial layers from the
pretrained checkpoint in forward patching, or from the SFT
checkpoint in reverse patching.

\subsection{Checkpoint Selection and Evaluation Protocol}
\label{app:patching_protocol}

For all nine model--learning-rate configurations, we patch checkpoints
$t=1,\ldots,10$ to examine early functional changes. Every checkpoint
is evaluated at all specified boundaries in both directions, together
with the unpatched pretrained and SFT baselines, using the dialect
test set and scoring protocol in Appendix~\ref{app:evaluation}.

All ten curves are displayed. We highlight steps 1 and 10 and an
intermediate step selected from the unpatched CER trajectory:
\[
t_{\mathrm{peak}}(m,\eta)
=\underset{t\in\{2,\ldots,9\}}{\arg\max}
\mathrm{CER}(M_{m,\eta}^{(t)}).
\]
Selection precedes inspection of the patching curves and is shared
across boundaries and directions. The selected steps are
\[
\begin{array}{lccc}
\toprule
 & 10^{-3} & 10^{-4} & 10^{-5} \\
\midrule
\text{Qwen3-ASR} & 6 & 3 & 4 \\
\text{FireRed-LLM} & 7 & 5 & 5 \\
\text{MiMo-Audio} & 8 & 6 & 5 \\
\bottomrule
\end{array}
\]
The remaining curves are shown in lighter colors.

\subsection{Additional Results Across Learning Rates}
\label{app:patching}

\begin{figure*}[t]
    \centering
    \includegraphics[width=\textwidth,height=0.27\textheight,
                     keepaspectratio]{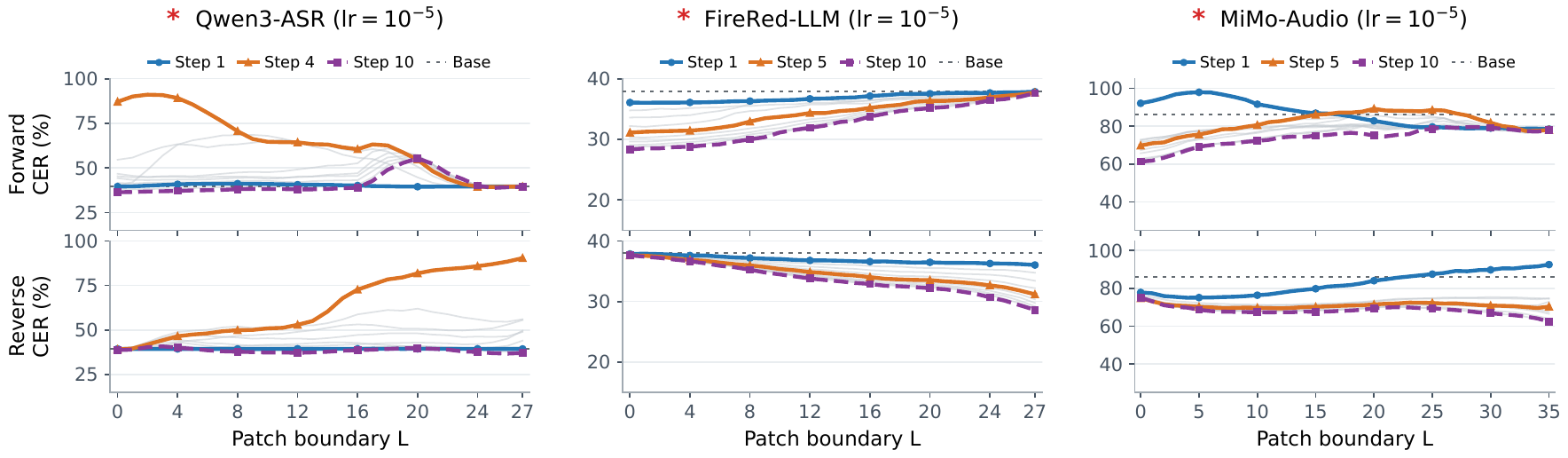}

    \vspace{0.5em}
    \includegraphics[width=\textwidth,height=0.27\textheight,
                     keepaspectratio]{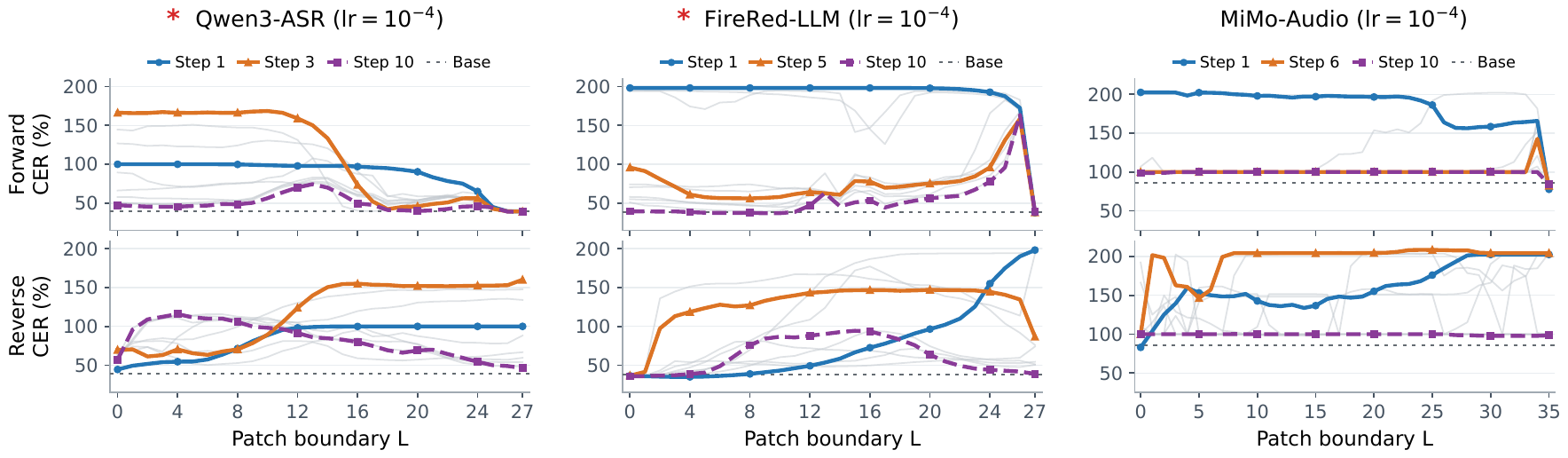}

    \vspace{0.5em}
    \includegraphics[width=\textwidth,height=0.27\textheight,
                     keepaspectratio]{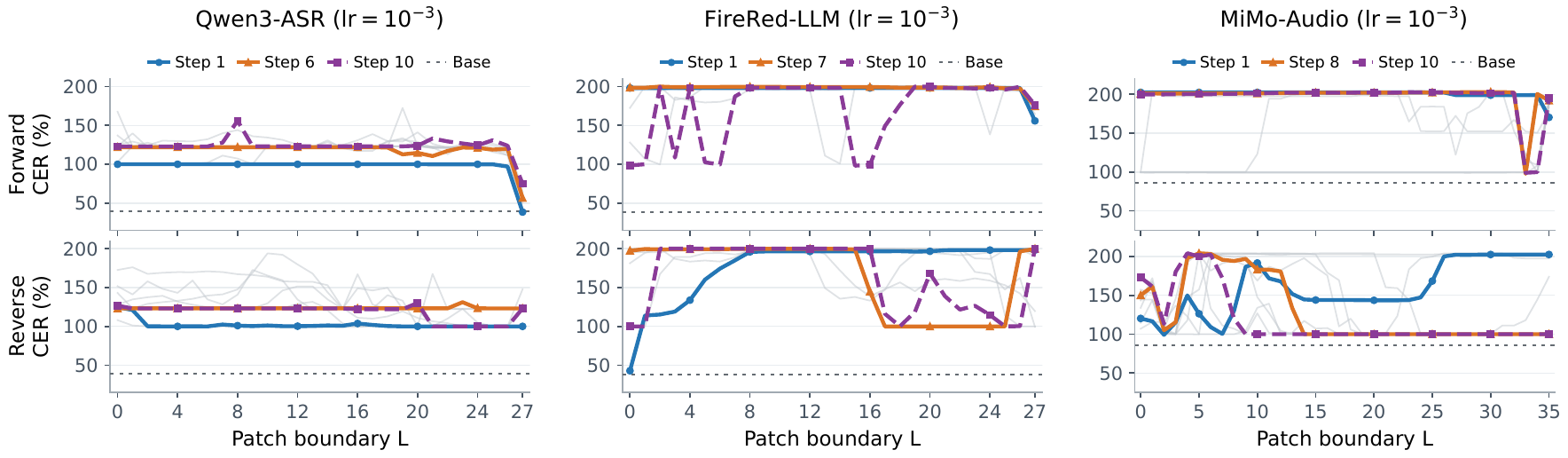}

    \caption{Bidirectional layer patching at learning rates
    $10^{-5}$ (top), $10^{-4}$ (middle), and $10^{-3}$ (bottom).
    Red stars beside model names indicate runs with effective adaptation.
    Within each plot, the upper row shows forward patching and
    the lower row shows reverse patching. Colored curves highlight
    steps 1, 10, and $t_{\mathrm{peak}}$; gray curves show the
    remaining early checkpoints.}
    \label{fig:patching_all_lr}
\end{figure*}

Figure~\ref{fig:patching_all_lr} presents the complete
bidirectional patching results for all three models at
$10^{-5}$, $10^{-4}$, and $10^{-3}$, extending the
collapse-and-repair cases discussed in Fig.~\ref{fig:patching}.
At $10^{-5}$, the layer-patching profiles are generally smoother,
especially for FireRed-LLM. As the learning rate increases, the
checkpoint with the largest disruption within the first ten updates
shows progressively stronger cross-checkpoint incompatibility across
network depth. At $10^{-3}$, this mismatch becomes broadly distributed
across the network, with no consistent layer region capable of
recovering recognition. This behavior is consistent with a transition
from localized and recoverable functional disruption to near-global
loss of usable pretrained computation, accompanied by persistent
recognition failure and ineffective convergence.

\section{Gradient Similarity Analysis}
\label{app:gradient_similarity}

\subsection{Gradient Extraction}
\label{app:gradient_extraction}

We measure dialect-specific gradients online for all nine SFT
configurations. At training step $t$, the gradient for dialect $i$ is
\[
g_{t,i}
=\nabla_{\theta_t}\mathcal L(\mathcal B_{t,i};\theta_t),
\qquad i\in\{1,\ldots,5\}.
\]
Here, $t$ indexes optimizer updates, $i$ indexes the five dialects,
$\mathcal B_{t,i}$ is the batch of dialect-$i$ utterances used at
step $t$, and $\theta_t$ is the selected decoder parameter state
immediately before that update. The loss $\mathcal L$ is the cross-entropy (CE) loss computed with teacher forcing, and $g_{t,i}$ is its gradient with respect
to $\theta_t$. At a full training step, each $\mathcal B_{t,i}$
contains 128 utterances. All five dialect gradients at step $t$
are measured at the same parameter state.

For Qwen3-ASR and FireRed-LLM, the selected parameters comprise
all trainable text-decoder components: embeddings, Transformer
layers, final normalization, and the output head. For MiMo-Audio,
they comprise the global text decoder and LM head, excluding local
audio Transformers and audio heads. Frozen acoustic components
are excluded for all models.

We aggregate each dialect gradient across GPUs and accumulation
micro-steps, then concatenate its parameter gradients into a single
vector without layerwise normalization. These measurements are
taken before gradient clipping and optimizer transformations and
do not alter training. Within each model, learning-rate runs share
the data sampling procedure, dialect order, and random seed.
Because batches differ between steps, comparisons over time
reflect both parameter changes and sampling variation.

\subsection{Within-Step and Inter-Step Similarity}
\label{app:intra_step_similarity}
\label{app:inter_step_similarity}

For each nonzero dialect gradient, we define its unit vector as
\[
\bar g_{t,i}=\frac{g_{t,i}}{\|g_{t,i}\|_2},
\]
We measure cross-dialect similarity within step $t$ and
same-dialect similarity between consecutive steps as
\[
s_{t,ij}=\bar g_{t,i}^{\top}\bar g_{t,j}
\quad(i<j),
\qquad
r_{t,i}=\bar g_{t,i}^{\top}\bar g_{t-1,i}
\quad(t\geq2).
\]
Here, $i$ and $j$ denote different dialects. Both dot products
are computed over all selected decoder parameters. Values near
$1$, $0$, and $-1$ indicate aligned, approximately orthogonal,
and opposing gradient directions, respectively.

With five dialects, each step yields ten cross-dialect values,
and each pair of consecutive steps yields five same-dialect
values. In particular, $r_{t,i}$ compares gradients measured
before optimizer updates $t-1$ and $t$.

The within-step mean reported in the main text is
\[
S_{\mathrm{intra}}(t)
=\frac{1}{10}\sum_{i<j}s_{t,ij}.
\]
The values $r_{t,i}$ give the dialect-resolved form of the
inter-step similarity $S_{\mathrm{inter}}$. For visualization, we use the medians specified below.

\subsection{Summary Statistics and Visualization}
\label{app:gradient_similarity_statistics}

In Fig.~\ref{fig:gradient_similarity}, gray curves show the ten
dialect pairs or five individual dialects. Colored curves show
their pointwise median, and shaded bands span the 25th--75th
percentiles. These bands describe variation across pairs or
dialects. Cross-dialect measurements
cover steps 1--391 and inter-step measurements cover steps 2--391. The zero line marks
orthogonality.

\section{Gradient Spectral Analysis}
\label{app:gradient_spectrum}

\subsection{Gradient Matrix Construction}
\label{app:gradient_matrix}

Using the gradients defined in Appendix~\ref{app:gradient_extraction},
we stack the five dialect gradients as rows of a matrix:
\[
G_{\mathrm{raw}}^{(t)}
=
\begin{bmatrix}
g_{t,1}^{\top}\\
\vdots\\
g_{t,K}^{\top}
\end{bmatrix}
\in\mathbb R^{K\times P},
\qquad K=5.
\]
Here, $t$ indexes training steps, $g_{t,i}\in\mathbb R^P$ is the
gradient for dialect $i$ measured before update $t$, $K$ is the
number of dialects, and $P$ is the number of selected decoder
parameters. We do not normalize the rows, so the spectrum captures
both gradient directions and relative magnitudes. This is the
raw-gradient matrix used in Section~\ref{sec:gradient_measurements}.

\subsection{Singular Values and Spectral Energy}
\label{app:svd}

We compute the singular values of $G_{\mathrm{raw}}^{(t)}$ through
its $K\times K$ Gram matrix:
\[
H^{(t)}
=G_{\mathrm{raw}}^{(t)}(G_{\mathrm{raw}}^{(t)})^\top,
\qquad
H_{ij}^{(t)}=g_{t,i}^{\top}g_{t,j}.
\]
Here, $i$ and $j$ index dialects, and $H_{ij}^{(t)}$ is the
inner product of their gradients at step $t$. Let
$\lambda_k^{(t)}$ be the $k$-th largest eigenvalue of $H^{(t)}$,
where $k=1,\ldots,K$. The corresponding singular value is
\[
\sigma_k^{(t)}=\sqrt{\max(\lambda_k^{(t)},0)}.
\]
The maximum with zero removes small negative eigenvalues caused
by numerical round-off. We retain all $K$ singular values,
including near-zero values.

The spectral energy share of component $k$ is
\[
p_k^{(t)}
=
\frac{(\sigma_k^{(t)})^2}
     {\sum_{\ell=1}^{K}(\sigma_\ell^{(t)})^2},
\qquad
\sum_{k=1}^{K}p_k^{(t)}=1.
\]
The denominator is the total squared gradient norm,
$\|G_{\mathrm{raw}}^{(t)}\|_F^2$, where $\|\cdot\|_F$ denotes
the Frobenius norm. Figure~\ref{fig:gradient_spectrum} plots
$p_k^{(t)}$ in descending energy order at each step.

\subsection{Additional Gradient Spectral Analysis}
\label{app:spectral}

The participation ratio reported in the main text
is 
\[
D_{\mathrm{eff}}^{(t)}
=
\frac{1}
{\sum_{k=1}^{K}\left(p_k^{(t)}\right)^2}.
\]
With five sampled gradients, $1\leq D_{\mathrm{eff}}^{(t)}\leq5$:
values near 1 indicate that the spectral energy is concentrated in
a small number of components, whereas larger values indicate a more
distributed spectrum. This quantity is an effective dimensionality,
rather than an integer matrix rank.

\begin{figure}[h]
    \centering
    \includegraphics[width=\linewidth]{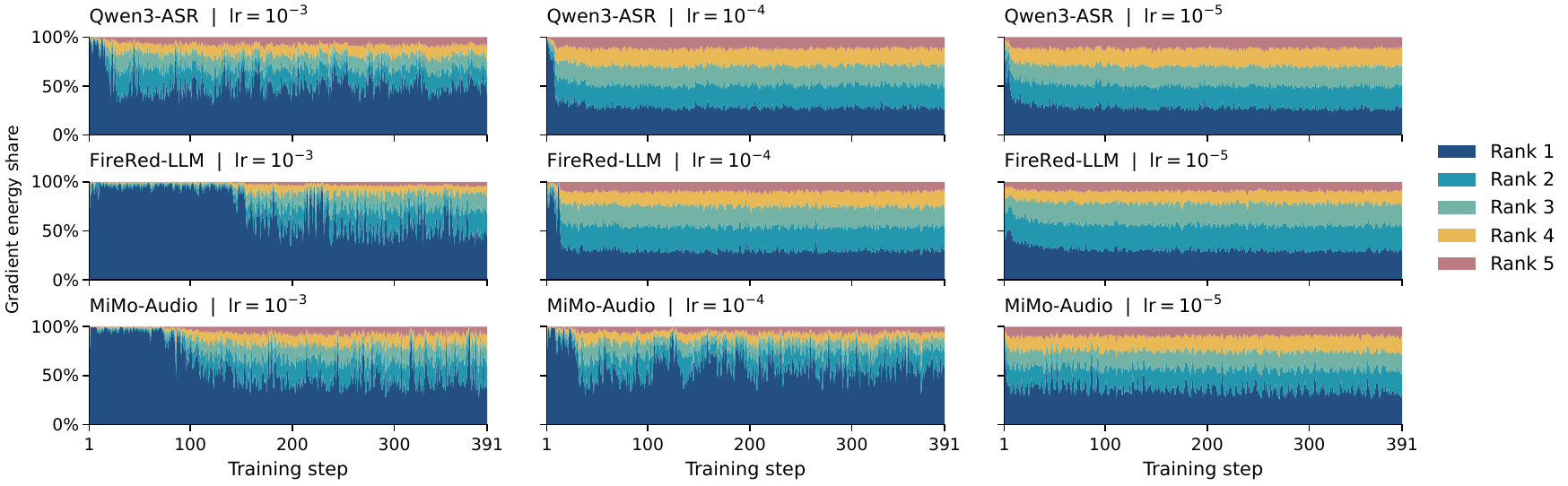}
    \caption{Evolution of gradient spectral energy during SFT across three models and learning rates. Stacked areas show the energy share of each singular component, $p_k=\sigma_k^2/\sum_j\sigma_j^2$, ordered by descending singular value at each training step.}
    \label{fig:gradient_spectrum}
\end{figure}

Figure~\ref{fig:gradient_spectrum} shows that successful runs
need not follow the same early spectral trajectory. FireRed-LLM
at $10^{-5}$ has a relatively balanced energy distribution from
the early steps, whereas Qwen3-ASR at $10^{-5}$ and both models
at $10^{-4}$ redistribute energy across components before
settling into relatively stable patterns. In the failed
$10^{-3}$ runs, energy remains concentrated for extended periods
or its distribution continues to fluctuate substantially.
The full spectra also show that similar values of
$D_{\mathrm{eff}}^{(t)}$ can correspond to different allocations
of energy among singular components.

\section{AdamW Update Transfer: Implementation}
\label{app:update_transfer}

We provide the implementation details of the AdamW update-transfer
intervention described in Section~\ref{sec:intervention}. The
intervention is applied to Qwen3-ASR at learning rates of $10^{-4}$
and $10^{-5}$ and to FireRed-LLM at $10^{-4}$, with all three
configurations reported in Fig.~\ref{fig:adamw_transfer}. All runs
use seed 5230 and a constant learning rate without warm-up.

Throughout this section, $s$ denotes the zero-based global batch-step
index used in Fig.~\ref{fig:adamw_transfer}. The teacher's first ten
updates therefore correspond to $s=0,\ldots,9$, while the pretrained
checkpoint precedes $s=0$. Elsewhere in the paper, $M^{(0)}$ denotes
the pretrained checkpoint and $t=1,\ldots,391$ counts completed
optimizer updates.

\subsection{Teacher and Student Initialization}

Teacher and student are initialized from the same pretrained checkpoint
and optimize the text-decoder parameters while keeping the acoustic
components frozen. The teacher follows ordinary SFT throughout the
experiment. The student remains at the pretrained model parameters
during global steps $s=0,\ldots,9$ and begins updating at
$s_{\mathrm{start}}=10$.

Immediately after the teacher completes its update at $s=9$, we copy
the teacher's AdamW optimizer state to the student without transferring
any model parameters. For each trainable parameter, this includes the
first-moment estimate, the second-moment estimate, and the optimizer step counter, which equals 10
after the first ten teacher updates. Consequently, when the student is
activated at $s=10$, its model parameters remain exactly at the
pretrained initialization, while its AdamW state reflects the teacher's
first ten optimization steps.

\subsection{Update Mixing and Schedule}
\label{app:gradient_mixing}

From global step $s=10$ onward, teacher and student process the same
training batches of 640 utterances across eight GPUs. 

Each model independently accumulates gradients, clips the global
gradient norm to 1.0, and computes its complete AdamW parameter
displacement, denoted by $u_s^T$ for the teacher and $u_s^S$ for the
student. These displacements include the effects of the AdamW moment
estimates, bias correction, learning-rate scaling, and decoupled weight
decay. The teacher applies its ordinary update $u_s^T$, whereas the
student applies
\begin{equation}
\theta_{s+1}^{S}
=
\theta_s^{S}
+
a_s u_s^{S}
+
b_s u_s^{T}.
\end{equation}
Thus, the intervention mixes the complete AdamW parameter
displacements rather than the raw gradients. No projection or
update-norm matching is applied.

The mixing coefficients follow
\[
(a_s,b_s)=
\begin{cases}
(0,0), & 0\leq s<10,\\[3pt]
\left(
0.125\dfrac{s-10}{30},
0.125\dfrac{40-s}{30}
\right),
& 10\leq s\leq40,\\[8pt]
\left(
0.125+0.875\dfrac{s-40}{30},
0
\right),
& 40<s\leq70,\\[8pt]
(1,0), & s>70.
\end{cases}
\]

At the activation step $s=10$, the student therefore applies only a
scaled teacher update, with $(a_{10},b_{10})=(0,0.125)$. From
$s=10$ to $s=40$, teacher influence is gradually removed while the
student contribution increases to 0.125. From $s=40$ to $s=70$, the
teacher contribution remains zero and the student coefficient increases
from 0.125 to 1. After $s=70$, the student resumes ordinary SFT.

The student's AdamW moments continue to evolve throughout this process.
Only the parameter displacement applied at each step is mixed; the
student optimizer state is neither replaced nor reset after activation.

\subsection{Continued Training and Evaluation}
\label{app:transfer_evaluation}

Teacher and student continue training to the end of the same SFT epoch.
After $s=70$, the student follows ordinary AdamW optimization without
teacher influence or any optimizer-state reset. The uninterrupted
teacher serves as the matched baseline.

Checkpoints are indexed by the shared zero-based global data-stream
step $s$, rather than by the number of student updates since activation.
We save the pretrained checkpoint, checkpoints after every indexed
update from $s=0$ through $s=80$, and subsequently every ten steps
together with the final checkpoint. Validation loss is evaluated at
these checkpoints, while CER is computed offline using the common
evaluation protocol described in Appendix~\ref{app:evaluation}.

\section{Effect of Learning-Rate Warm-Up}
\label{app:warmup}

We examine whether learning-rate warm-up mitigates early CER
degradation in three configurations: FireRed-LLM at $10^{-4}$ and
Qwen3-ASR at $10^{-4}$ and $10^{-5}$. The two $10^{-4}$ configurations
exhibit pronounced early CER spikes in both domains; Qwen3-ASR at
$10^{-5}$ shows early Mandarin deterioration and a dialect CER plateau.
We exclude configurations
with smooth adaptation or persistent optimization failure,
as this experiment focuses on transient recognition degradation.

For each configuration, we compare SFT without warm-up against
SFT with a warm-up ratio of 0.02. During warm-up, the learning
rate increases linearly to the target value and remains constant
thereafter. All other training settings are unchanged. We compare test-CER
trajectories on Mandarin and dialect speech.

\begin{figure}[h]
    \centering
    \includegraphics[width=\linewidth]{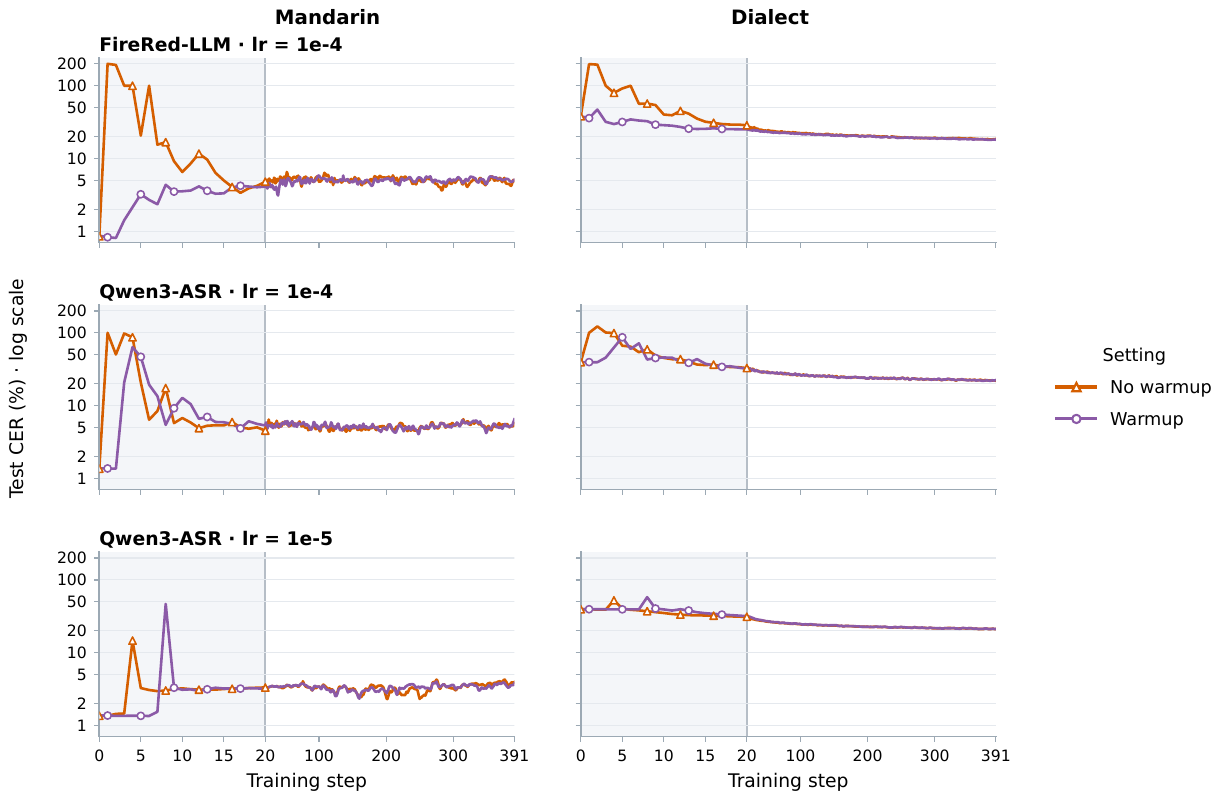}
    \caption{
    Test CER with and without learning-rate warm-up
    (warm-up ratio: 0.02).
    Rows show FireRed-LLM at $10^{-4}$, Qwen3-ASR at $10^{-4}$,
    and Qwen3-ASR at $10^{-5}$, respectively.
    Columns show Mandarin (left) and dialect speech (right).
    Orange and purple curves denote runs without and with
    warm-up, respectively.
    The vertical axis is logarithmic; the horizontal axis uses
    different linear scales before and after step~20 to
    highlight early training.
    }
    \label{fig:warmup_comparison}
\end{figure}

Figure~\ref{fig:warmup_comparison} shows that the effect of
warm-up depends on the model and learning rate.
For FireRed-LLM at $10^{-4}$ (top row), warm-up substantially
suppresses the early CER spikes on both test domains.
For Qwen3-ASR at $10^{-4}$ (middle row), it delays and reduces
the peaks but does not eliminate them.
At $10^{-5}$ (bottom row), it delays and increases the magnitude of
early Qwen3-ASR Mandarin deterioration, while dialect adaptation
remains delayed.

Despite these differences, runs with and without warm-up
reach similar final CER on both test domains.
Thus, a warm-up ratio of 0.02 does not consistently prevent
early CER spikes in the tested configurations, nor does it
substantially improve final Mandarin retention.

\section{Representative Recognition Error Cases}
\label{sec:recognition_error_cases}

\begin{table*}[h]
\centering
\small
\caption{Representative dialect ASR predictions during early recognition
collapse in Qwen3-ASR at step 3 with a learning rate of $10^{-4}$.
This run subsequently recovers and continues effective training.}
\label{tab:repetitive_prediction_cases}
\begin{tabular}{p{0.12\textwidth} p{0.73\textwidth} p{0.08\textwidth}}
\toprule
Dialect & Transcript & CER \\
\midrule

Hakka
& \textbf{Reference:}
  \begin{CJK*}{UTF8}{gbsn}把跑车开转屋卡酿边料\end{CJK*}
  \newline
  \textbf{Hypothesis:}
  \begin{CJK*}{UTF8}{gbsn}今天只有今天只有只有今天今天今天今天今天\end{CJK*}
& 200.00\% \\

\addlinespace
Changsha
& \textbf{Reference:}
  \begin{CJK*}{UTF8}{gbsn}多好多好那都是假的\end{CJK*}
  \newline
  \textbf{Hypothesis:}
  \begin{CJK*}{UTF8}{gbsn}今天只有明天只有明天明天明天明天明天\end{CJK*}
& 177.78\% \\

\addlinespace
Nanchang
& \textbf{Reference:}
  \begin{CJK*}{UTF8}{gbsn}我不想拉个脸对你\end{CJK*}
  \newline
  \textbf{Hypothesis:}
  \begin{CJK*}{UTF8}{gbsn}今天只有今天只有今天只有今天今天\end{CJK*}
& 200.00\% \\

\bottomrule
\end{tabular}
\end{table*}

\begin{table*}[h]
\centering
\small
\caption{Representative dialect ASR predictions during persistent
optimization failure in Qwen3-ASR at step 150 with a learning rate
of $10^{-3}$.}
\label{tab:short_repetitive_cases}
\begin{tabular}{p{0.12\textwidth} p{0.72\textwidth} p{0.09\textwidth}}
\toprule
Dialect & Transcript & CER \\
\midrule

Hakka
& \textbf{Reference:}
  \begin{CJK*}{UTF8}{gbsn}康熙帝有几多只妹哩\end{CJK*}
  \newline
  \textbf{Hypothesis:}
  \begin{CJK*}{UTF8}{gbsn}捱唔唔唔唔唔唔\end{CJK*}
& 100.00\% \\

\addlinespace
Nanchang
& \textbf{Reference:}
  \begin{CJK*}{UTF8}{gbsn}不是飞机上咯啊\end{CJK*}
  \newline
  \textbf{Hypothesis:}
  \begin{CJK*}{UTF8}{gbsn}我什哩咯\end{CJK*}
& 85.71\% \\

\addlinespace
Ningbo
& \textbf{Reference:}
  \begin{CJK*}{UTF8}{gbsn}事体发生咯第二日记者上门去查看门锁的\end{CJK*}
  \newline
  \textbf{Hypothesis:}
  \begin{CJK*}{UTF8}{gbsn}阿拉咯咯咯咯咯咯咯咯咯咯咯\end{CJK*}
& 94.44\% \\

\addlinespace
Changsha
& \textbf{Reference:}
  \begin{CJK*}{UTF8}{gbsn}三人同是常长客运专线联营车队抱团自封的车队管理人员\end{CJK*}
  \newline
  \textbf{Hypothesis:}
  \begin{CJK*}{UTF8}{gbsn}22222222年的2年的的\end{CJK*}
& 96.00\% \\

\addlinespace
Minnan
& \textbf{Reference:}
  \begin{CJK*}{UTF8}{gbsn}耐克的鞋子但是不知道是什么款式的\end{CJK*}
  \newline
  \textbf{Hypothesis:}
  \begin{CJK*}{UTF8}{gbsn}你我你你你你什\end{CJK*}
& 93.75\% \\

\bottomrule
\end{tabular}
\end{table*}

\end{document}